# EXPERIMENTAL RESEARCH DATA QUALITY IN MATERIALS SCIENCE

Thorsten Wuest[1], Rainer Tinscher[2], Robert Porzel[3] and Klaus-Dieter Thoben[1,4]

[1] Department ICT applications for production, BIBA - Bremer Institut für Produktion und Logistik GmbH, Hochschulring 20, 28359 Bremen, Germany
[2] Stiftung Institut für Werkstofftechnik, Division Materials Science, Badgasteiner Straße 3, 28359 Bremen, Germany
[3] University of Bremen, Research Group Digital Media, Bibliothekstraße 1, 28359 Bremen, Germany
[2] University of Bremen, Institute for Integrated Product Development, Bibliothekstraße 1, 28359 Bremen, Germany

*ABSTRACT*

*In materials sciences, a large amount of research data is generated through a broad spectrum of different experiments. As of today, experimental research data including meta-data in materials science is often stored decentralized by the researcher(s) conducting the experiments without generally accepted standards on what and how to store data. The conducted research and experiments often involve a considerable investment from public funding agencies that desire the results to be made available in order to increase their impact. In order to achieve the goal of citable and (openly) accessible materials science experimental research data in the future, not only an adequate infrastructure needs to be established but the question of how to measure the quality of the experimental research data also to be addressed. In this publication, the authors identify requirements and challenges towards a systematic methodology to measure experimental research data quality prior to publication and derive different approaches on that basis. These methods are critically discussed and assessed by their contribution and limitations towards the set goals. Concluding, a combination of selected methods is presented as a systematic, functional and practical quality measurement and assurance approach for experimental research data in materials science with the goal of supporting the accessibility and dissemination of existing data sets.*

## 1. INTRODUCTION

The topic of data quality gains importance due to various factors such the rapid development of the so-called "big data" movement (Lohr, 2012; Aggrawal, 2013; Küll, 2013; Lycett, 2013; Kwon, Lee, & Shin, 2014), increasing automation of manufacturing systems, and developments in Information and Communication Technology (ICT). With this development, practitioners and academics in various domains (Ozmen-Ertekin & Ozbay, 2012) have to face the challenge of ensuring the quality of generated or captured data. In academia, where researchers base their conclusions on research data, the quality of the data has a tremendous impact on the results. Especially when sharing information, the importance of inherited data quality can not be rated high enough (Hartono, Li, Na & Simpson, 2010). However, Büttner et al., state in their handbook on research data management, that the absence of a chapter "research data quality" is highlighting the need to begin exploring this issue (Büttner, Hobohm & Müller, 2011).

There are various publications and practical handbooks about data quality issues (e.g., Pierce, 2004; Evans, 2006; Naumann, 2007; Büttner, Hobohm & Müller, 2011). However, most of them are very specific for a certain type of data sets or domain. However, research data has certain requirements and differentiates itself from other data sets. As scientific data are in many cases very heterogeneous and can even be based on unique experimental designs, it is quite difficult to assess and measure data quality. This is where this publication seeks to make a contribution by first presenting the specifications of experimental research data per se and then focus on the domain of materials science, followed by a section elaborating on data quality and the special requirements of experimental research data with respect to data quality. Based on the findings, the fourth section will present approaches that support the quality assurance of experimental research data and map them in a matrix to the requirements of the research domain of materials science. In the fifth section, a methodology to assure experimental research data quality in materials science is presented and critically discussed. The final section provides a short conclusion and an outlook on this important topic.

## 2. STATE OF THE ART IN RESEARCH DATA IN MATERIALS SCIENCE

In this section, a brief recapitulation of the definition and meaning of research data itself and its significance within the materials science domain is presented. Furthermore, characteristics of experimental research data in materials science are derived as a foundation for the presented research in the following sections.

### 2.1. Experimental research data

Most research domains share similar problems when looking at the research data they produce. In most cases there are no standards in form of format style or documentation. Therefore, the research data is generated or saved in various formats, which are most of the time not consistent semantically and meta-data is provided incompletely (Neuroth et al. 2009). This holds true for most empirical research fields. The research domain of materials science can be located in the group of research domains, which conduct empirical experiments in order to obtain new findings. The absence of standards in format and incomplete provision of meta-data hinders the reuse and thus the continuous exploitation of experimental research data to a large extent. Be it reuse by the same institute/researchers or another, each one requires a certain quality of the data and its documentation, thus data management in order to be able to work with the research data. In order to ensure the sustainability of the data and the maximum impact of mostly expensive experiments conducted to generate it, the data quality of the research data has to be secured.

Experimental research data is often 'generated' by informal processes, which are adapted based on the research question (Neuroth et al., 2009). Major funding organizations put an increased focus on research data management over the last years. In 2009 the German Research Foundation (DFG) published a number of recommendations concerning the handling of research data. Within the first recommendation the term research data is broadly defined as all data, which is generated during study of sources, experiments, measurements, surveys and polls. An important aspect is that research data is seen as the foundation of scientific publications (DFG, 2009). Within this definition, every institution has the freedom to choose to what extent and what state of aggregation the research data is to be saved. However, the second recommendation puts forward the need for domain specific organizational standards for the handling of research data. It is argued, that the diverse nature of the different research domains is reflected in the data and thus the individual requirements towards a sufficient data management (DFG, 2009).

Monteagudo (1995) described the increasing challenges for the management of experimental research data put forward by e.g. the developments in instrumentation. The amount of data increases due to automatic measurements by sensors and/or computers (Scott, et al., 2013). The contextual information is often not considered anymore (Monteagudo, 1995) and it is almost impossible to access this contextual information at a later stage of the experiment. This is a key factor of success for any platform handling research data and any future infrastructure has to be measured by its contribution to this. The next sub-section further analyses the research data in materials science.

## 2.2. Materials science research data

Data management and corresponding infrastructure is to a large extent domain specific (Büttner, Hobohm & Müller, 2011). In materials science, as mentioned before, research data is the results of a wide variation of different experiments and observations. Occurrences in and on materials, e.g. happening during an experiment, represent the time dimension of an observation and describe internal and external changes of the material sample (Ilschner & Singer, 2010). Experiments and series of experiments test the properties of materials under a variation of circumstances. In order to perform the experiments, major investments are often necessary. Depending on the experiment, the material and the other factors, e.g., test-equipment and pre-treatment of the material, the generation of a research data set in the materials sciences can be very costly and time consuming. One lever to realize monetary and time savings within material testing is seen in the increasing amount of simulation activities in the field of materials science during the last 20 years. On the other hand, solid simulation models need to be feed with reliable experimental research data function probably and provide dependable results.

Most of the experimental research data is based on measurements and observations and provide generally a good foundation for processing in ICT systems/infrastructures. In many cases the data is based on standardized tests and experiments. To give an impression on the diversity and heterogeneity of the research data, it is important to understand the wide range of different experiments, which are conducted within the materials sciences. This can be for example tensile tests, fatigue tests, bend or hardness tests all described by standards like DIN, EN, ASTM, ISO etc. specified for different materials, e.g. metals, polymers, carbon fiber reinforced plastics, etc. These standards however to not specify how and to what extent the resulting experimental research data shall be saved or archived in an information infrastructure. Nor do they specify how to measure the quality of the resulting experimental research data. Hence, there is a lack of guidance concerning a standardized handling of experimental research data in the materials sciences.

However, it has to be noted, that each of the experiments can be altered to some extent, changing the results significantly or just slightly. This is an important part of the ongoing research activity in the materials science domain and therefore has to be considered when talking about the handling of research data. The adaption of existing or definition of new experiments plus the large amount of different materials increases the complexity continuously.

Nevertheless, it is important to create an environment where the generated research data is as well prepared in terms of data quality as possible for future continuous exploitation and to ensure a quick search and find for older data. Not only because the generation involved large investments both in time and money, but also because it allows for a later verification of the conducted research results as well as the employment of data mining and knowledge discovery approaches.

## 3. QUALITY OF EXPERIMENTAL RESEARCH DATA IN MATERIALS SCIENCE

Data quality is a one of the major factors when it comes to potential reuse of existing data not only in materials science but also in most research domains. Before looking into methods and tools designed for measuring and thus ensuring data quality, a common understanding of quality, data quality and, more specifically, data quality of experimental research data in materials science will be established in this section.

### 3.1. Quality

Quality has been a focus area of various engineering domains like manufacturing for several decades and the market success of companies successful in utilizing their understanding of quality and customer requirements highlight the importance of quality. De Weck, Ross & Rhodes (2012) found in their recent study on system lifecycle properties ('Ilities') that quality is and was the most dominant 'ility' of engineering systems for over a century, rated higher than e.g., reliability and safety (De Weck, Ross & Rhodes, 2012).

In this paper the term quality is understood as "the degree to which a set of inherent characteristics fulfills requirements" (DIN EN ISO 9001:2008). Requirement within this context is defined as the "need or expectation that is stated, generally implied or obligatory" (DIN EN ISO 9001:2008). According to this definition, quality depends on the fulfillment of requirements. The fulfillment of these requirements depends on the planning of processes (commands) and the execution of processes (executions) (see Figure 1) (Jacob & Petrick, 2007). Quality of the final product is regarded as achieved to a higher degree when more of the original customer requirements match with the achieved final product characteristics (Sitek, 2012).

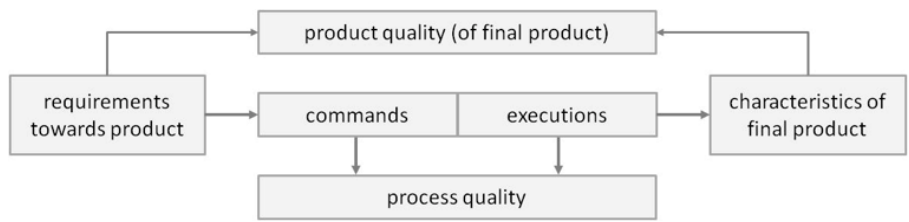

Figure 1. Elements of quality (adapted from Jacob & Petrick, 2007; Sitek, 2012)

It has to be considered that a product can inherit different qualities, the sum of these qualities like e.g., security, workmanship, durability or stable value represent the final product quality (Kamiske & Brauer, 2008). Several of these qualities can be represented by material properties. In the next section, the meaning of quality of data, so-called data quality, is explained.

### 3.2. Data quality

Data quality is a multi-dimensional concept (Pipino, Lee & Wang, 2002). Data and information quality is usually defined in terms of contribution to the objectives of the end-user (Helfert, 2002). It can be additionally described as the adequacy for the relevant data processing application (Naumann, 2007). Poor data quality can be a major cause for damages and losses on organizational processes (Storey et al., 2012). To avoid the damages and losses data quality problems and solutions should be considered as early as possible, best at the design stage of the information sys-tem (Storey et al., 2012).

Pipino, Lee & Wang, 2002 lists under the data quality dimensions the following attributes: accessibility, appropriate amount of data, believability, completeness, concise representation, consistent representation, ease of manipulation, free-of-error, interpretability, objectivity, relevancy, reputation, security, timeliness, understandability, value-added (Pipino, Lee & Wang, 2002). Data quality can thus be described as a set of quality characteristics (Naumann, 2007). Many of the listed attributes contribute to the overall data quality, as tested by a third-peer.

The starting point for consideration of data quality is the user-oriented quality concept. Helfert divides data quality in design and execution quality. The fulfillment of end-user requirements and specifications can be met through a choice of properties in the data design. Design quality refers as such to the collection of specific quality requirements from the user's perspective. Execution quality includes compliance with the specifications (Helfert, 2002). Helfert's basic data quality criteria are: correctness, completeness, consistency and timeliness (Helfert, 2002). Data quality criteria developed by English are: data standards, data definitions and information architecture (English, 1999). These criteria can be understood as the access capability, timeliness and interpretability of the data and the data system. Data quality as understood by Wang and Strong can be divided into internal data quality, contextual data quality, presentation, and access quality (Wang & Strong, 1996). Wang and Strong focus on user-related data quality - interpretability, usefulness, credibility, time reference, and availability have been rated as important criteria (Wang & Strong, 1996; Helfert, 2002). Jarke, Jeusfeld, Quix & Vassilidis's data quality criteria are: completeness, credibility, accuracy, consistency, and interpretability (Jarke et al., 1999). A poll conducted by Helfert delivers additional quality criteria important to organizations: clearly defined data descriptions, formal data syntax, delivery times (for data), and specific information about selected properties of the data, e.g., number of errors (Helfert, 2002).

Rohweder et al. describe data quality as the degree the characteristics suffice the requirements on the data product. The requirements for the data are determined through particular decisions and goals set on data quality. Rohweder et al. define data quality with the help of 15 IQ (Information Quality) dimensions (Rohweder, Kasten, Malzahn, Piro & Schmid, 2011). These can be applied to e.g., master data, to assess if the data is useful or not acceptable. The IQ dimensions have been divided into four categories and form a regulatory concept for data quality (Rohweder et al., 2011). The 15 IQ dimensions are as follows:

*System support (e.g., user interface)*

- Accessibility: accessible & easy to access.
- Ease of manipulation: easy to use & to change.

*Inherent (content examination)*

- Reputation: data source & processing highly trustworthy.
- Free of error: error free and consistent with reality.
- Objectivity: strictly objective and value-free.
- Believability: reinforced with quality standards, etc.

*Representation (overall presentation, e.g., the form of statistics)*

- Understandability: ability of users to directly understand & use information.
- Concise representation: clear, saved in appropriate & understandable format.
- Consistent representation: uniform & held in consecutive & equal manner.
- Interpretability: understandable in same, technically correct manner.

*Purpose-dependant (data use in the processes)*

- Timeliness: actual properties of data (described) accurately & up-to-date.
- Value-added: usage leads to quantifiable increase in monetary cost function.
- Completeness: no missing information contained.
- Appropriate amount of data: amount meets requirements set on data.
- Relevancy: provides all necessary information for user.

Overall, it is accepted, that all IQ dimensions should exhibit a high or at least sufficient quality for an information system to be functional (Rohweder et al., 2011). The following elaboration of data quality of experimental research data in materials science and the development of suitable measures is partly based upon the IQ dimensions presented in this section. Furthermore, the chosen approach is evaluated taking the requirements derived from the IQ dimensions into account.

### 3.3. Quality of experimental research data in materials science

The previous section presented a domain-independent elaboration of data and information quality. In the following section, the applicability of the described dimensions and specific characteristics of the materials science domain, which need to be considered, are described.

The understanding of experimental research data quality in materials science complies with the before established understanding of information and data quality and its 15 dimensions. There are however certain aspects specific to the domain of materials science which have to be highlighted. For once, data quality of published data sets is very important. This is of course also true for other domains, but as in materials science basic parameters for later design and construction of critical product/parts may be set, there is an increased emphasis on this fact. Inaccurate data cannot only lead to wrong research results but endanger human life through e.g., failure of critical parts of a plane due to wrong dimensioning based on the inaccurate data available for a certain material. This can lead directly to the thread of large liability claims. Also in the materials science domain, research is often industry driven or at least involves industry, which is not only relevant by differing quality standards and sharing cultures but also concerning possible confidentiality agreements. This often limits the freedom of handling and processing the data and thus may limit the potential leverage of quality assurance approaches.

A more technical challenge of materials science research data towards quality is the existence of various data format (e.g., photos, graphs, plots, data all analog and/or digital) the data can be utilizing. Often combinations of the above-mentioned formats are represented in the resulting data set of an experiment. Other factors like e.g., the large number of tests, process parameters, quality of test equipment, available standards (e.g., DIN, ISO, EN, ASTM), etc. add to the inherit complexity and the diverse nature and high-dimensionality of the resulting data and meta-data. Finally the scientists often set own quality levels in accordance to the specific targets of the individual project. E.g. tensile tests performed in order to be used as a basis for numerical simulations require a higher data- and test-quality compared to tensile tests performed within a daily pre-delivery check of incoming basic materials. Nevertheless, both data-types need to fulfill specific quality requirements but there is no standardized scale to set them into a relation. In a database, both data sets might have a "green traffic light" for data quality but on different quality levels which cannot be compared. Information about the target quality within the meta data of a data set is therefore needed for making it usable for external third party users.

In the following section, different approaches to measure experimental research data quality are developed on the basis of a requirements analysis of the previously presented state of the art and an analysis of generally accepted and often-occurring data errors and challenges for data quality.

## 4. APPROACHES TO MEASURE DATA QUALITY IN MATERIALS SCIENCE

In this section, current requirements towards a systematic measurement of experimental research data quality are presented. This is fundamental for the subsequent development of practical approaches to measure research data quality in materials science. After the approaches are theoretically developed, they are compared in an illustrative matrix by their impact on certain key aspects of future application within the materials science domain.

### 4.1. Requirements of tools to assure data quality

Data errors can be avoided most effectively and most sustainably at the moment of their emergence, e.g. throughout the manual data entry or automatic data collection (Naumann, 2007). A direct capture of the data from the source to an electronic device without human interference is the best way to minimize data input errors. The same stands true for media discontinuity which may lead to loss or alteration of data. When human interface or media discontinuity is unavoidable, input errors may occur and consequently de-grade data quality (Verma, 2012). To prevent quality degradation, a quality check can be performed in the moment of data delivery. The data can be further checked by end-users through the use of complaints-forms or other rating-systems, e.g. statistical methods (Helfert, 2002). In case of data sets from external sources, it is essential for the researcher/information-manager to deal consciously with the data and the data quality; it is crucial to mark the problematic data to be able to deal consciously with it (Naumann, 2007). The external party and the person responsible for integrating the data into the target system should be clear about the purpose of why the data are being collected (e.g., goal of the experiment), and it should be clearly stated (Verma, 2012).

The most common data quality issues are incorrect or missing values, duplicates, and errors in the recording process (Helfert, 2002; Winkler, 2003; Naumann, 2007; Verma, 2012). Errors in data cause errors in reports generated from the data, thus reinforcing the "garbage-in-garbage-out-effect" (Naumann, 2007).

Errors can be found within the schema and/or the data level of the data. The schema level describes the errors in the structural, semantic and schematic heterogeneity of the data characteristics. The data level includes value, unit, accuracy, and duplicates errors (Naumann, 2007). Duplicates, one of the most costly data errors (Naumann, 2007) can arise, e.g. due to typographical errors in the unique identifiers (e.g. the name of the researcher) (Winkler, 2003). Missing identifiers and contradictions in data indicate low quality (Winkler, 2003). These issues can be prevented with data quality ensuring practices, e.g. marking of problematic data, auto correction of format errors, manual correction of the data values, troubleshooting and coordination with the data suppliers, and organizational rules (Helfert, 2002). Furthermore file-linkage can be used to create "more complete" data (Winkler, 2003).

The most important and critical characteristics of high-quality data are consistency, completeness and correctness, timeliness, referential integrity and syntactical correctness. The traceability of data origin and documentation of discrepancies is also relevant (Helfert, 2002). Semantics and identifiability, as well as the precision of value ranges, granularity of data models, and technical aspects of data are less critical for the overall data quality (Helfert, 2002).

Data quality can be assessed by a third-peer. The assessment can either be task-independent, where no contextual knowledge is required, or task-dependent, with specific application context (Pipino, Lee & Wang, 2002). The data quality methodologies can be classified according to various criteria (Batini & Scannapieco, 2006):

- Data-driven vs. process-driven
- Measurement vs. improvement (assessment or improvement of the data quality)
- General-purpose vs. specific-purpose
- Intra-organizational vs. inter-organizational

Assessing data quality is an on-going process (Pipino, Lee & Wang, 2002). The most common phases of the process for assessing and improving the data quality are finding the cause of error, designing improvement solutions on data, establishing process control, managing improvement solutions, and the regular measurement of the improvements (Batini & Scannapieco, 2006). Thus, e.g. after the data quality has been assessed, the results of the assessment have to be compared, the discrepancies identified, and the root causes of discrepancies determined. Last but not least the necessary actions for improvement have to be determined and taken into action (Pipino, Lee & Wang, 2002).

Helfert specifies the following checks as important when testing: value ranges, data types, duplicates, referential integrity, comparison to reference data, plausibility, review by end users or customers (Helfert, 2002).

Pipino, Lee & Wang, 2002 suggests the use of data quality metrics, e.g. a questionnaire to measure data quality dimensions (Pipino, Lee & Wang, 2002). Auditors can also use the method of the control matrix to evaluate information integrity, and thus the reliability of the information product, the data quality (Pierce, 2004). Pierce suggests the use of the following data quality dimensions in the control matrix: accuracy, accessibility, consistency and timeliness (Pierce, 2004). The dimensions chosen by Pierce are similar to the aforementioned criteria. Further methods for quality evaluation are questionnaires, data set sampling (small control sets) and the use of data profiling tools (Naumann, 2007). If the data is not reliable it will lose its importance and cease to be trustworthy. The data should always be up-to-date and each data item clearly defined (Verma, 2012).

Data quality as such is an organizational issue. Each step in the organizational process, from data capture to processing, has an impact on the final quality of the data. The costs of data management have to be evaluated and the process responsibilities (data management responsibilities) assigned within the organization. Because the value and costs of data quality are not identical for the data users, Storey suggests decentralized information systems, so the departments that value data quality most are given the responsibility over it (Storey, 2012). Identifying key data quality characteristics, setting clear data quality goals and building incentive systems to reward individuals who add value to the data quality are important parts in an organizational architecture set to maintain high data quality (Storey, 2012).

Different requirements on the data lead to different solutions on maintaining the data quality (Rohweder et al., 2011). Information managers have to choose the appropriate methods to maintain data quality and integrate these practices into the organizational processes to achieve long-term benefits of data quality.

Based on the presented challenges and requirements and additional, underlying requirements towards a successful measurement of research data quality, the following aspects have to be considered before deciding on how data quality can be measured and data errors avoided:

- Access to and knowledge of data and meta-data
- Ability to prevent fraud
- Ability to assure the same or comparable (quality) standards
- Cost of quality assurance approach (e.g., cost experimental data set)
- Amount of time and effort needed (by personnel involved)
- Starting point of quality assurance / Influence during process of data generation
- Acceptance of quality assurance approach in the respective research community

These points have to be targeted by potential approaches in order to have a chance for successful implementation in a research data infrastructure. All are important, whereas some are essential, e.g., the acceptance in the research community is essential as without this trust, the available research data will not be able to utilize its full potential for high impact reuse.

### 4.2. Analysis of approaches to measure experimental research data quality

Based on the previously described common data quality challenges and requirements, in this section, six different approaches on how research data quality within an information infrastructure in materials science can be secured are discussed.

#### 4.2.1. Author responsible

The author (owner, in most cases the researcher) of a research data set is the first person with significant influence on the data quality, being responsible for the accuracy and completeness. The author (researcher) in this case is the person responsible for an experiment, not necessary the person conducting the actual physical work (technician). The author therefore has the best insight on the data available and can set the lever for what meta-data is to be included as well as additional information which might be of relevance for future reuse. On the other hand, the author might be biased as the experiment is most likely conducted having a certain objective. The author might be tempted to just include the data relevant for reaching this objective, whereas neglecting other information which might be beneficial for other utilization. Also the sole responsibility of the author might not be beneficial for the comparability of different data sets due to varying standards concerning data quality in different areas and/ or institutions, etc.

It can be assumed that the author has a high interest in the quality of the data as he is the first user in most cases. However, research from various domains indicates that some researchers might have second thoughts of sharing their data with others (Olsen & Downey, 2012). The reasons may be manifold, reaching for the fear of competition to the insecurity of their own results. This indicates the need for an additional/complementary approach at the very least supplementing the author's sole responsibility of data quality assurance. The additional time and effort and not to forget cost of ensuring data quality by the author is, even considering variations, depend on the already existing archiving mechanisms and the type of experiment/machine/ material used, relatively low when the integration in the data generation process is smooth.

#### 4.2.2. Internal review (e.g., advisor/supervisor)

Internal review, in this case by a direct superior (advisor/supervisor/project manager) not by a peer (this case is included in the following sub-section) is another possible approach to ensure research data quality. In this case, a party not directly involved in the data generation but connected to the same department/institution is checking the data for consistency, completeness and the other dimensions described beforehand. The principle is based on a four-eye governing principle common in many disciplines. The advantage is that the internal reviewer has access not only to the original data but also to the process owner and/or resources used to generate the data. Therefore eventual questions concerning the data quality can be answered with relative little effort. The additional cost and time required depends strongly on the organizational structure and the number of assigned tasks per reviewer but can be considered higher than a quality assurance solely depending on the author. The possible bias, a main concern with the author based quality assurance, might be reduced when applying this approach as the supervisor/advisor may have a certain distance and intrinsic interest in an as broad as possible applicability of the generated data. However, this largely depends on the implementation and the review processes.

Concerning the comparability of used standards, this might lead to a separation and clustering of the existing research data in 'trustworthy' and 'not trustworthy' institutions, similar to what is happening to some extent in the field of journal publication with the dawn of many 'for-profit' online journals with questionable references. It would be beneficial for future users to avoid this and integrate some commonly accepted, non-biased 'proof' or 'indicator' for research data quality.

**4.2.3. External review (peers)**

Peer review is commonly accepted by scholars as of assuring the accuracy of results in scientific publications. However, even though the blind/non-blind peer review process is accepted as the best-known way to ensure the quality of new publications in various research fields, in the area of data publications the applicability and benefit has to be critically questioned. Whereas a publication describes the methodology, data and results in a comprehensive way, a data set on the other hand and especially surrounding meta-data and/or information about the experimental environment (e.g., type of machine, day/night, temperature, etc.) represent information which is hard to assess without being involved in the process itself.

The time and effort external reviewers would have to invest in order to achieve the in-depth understanding of the data needed to achieve acceptable results in judging the research data quality realistically exceeds the requirements the formerly presented two approaches with direct access to the data source might experience. It is even questionable if e.g., a blind peer review is possible in the case of data publications, as the chance for the reviewer needing additional information from the author may be considerably higher. The time-to-publication will most likely increase in case of external peer review activities, which has two important effects: first, the research data is not directly available to the interested public and second, in case of questions from the reviewer to the author, the chance that this information can be provided decreases with the time passed since the data is generated.

However, as peer reviews are commonly accepted, it can be assumed, that the community will accept a similar approach for data publications as well. And, similar to the existing practice for regular publications, the direct cost are very limited, as the reviewers mostly contribute free of charge. The indirect cost however may be significant, given the time spend on the reviews by researchers. Regarding the comparability and standardization, the peer review approach faces similar challenges as others, depending on individual input.

**4.2.4. External review (independent 3rd party, e.g., classification society)**

The fourth approach presented in this paper describes an independent third party being responsible for the (external) review of research data sets. This is known from standardized certifications like e.g., ISO9001 Quality Management audits. Specialized and trusted agencies are responsible for a neutral and professional review. The major advantage of such an approach is the (assumed) lack of bias and the comparability/standardization of results. The acceptance in the community can be assumed as being high if the independence and professional behavior of the reviewing party can be assured. Concerning the access to knowledge and meta-data, the access can be considered less problematic than for a (blind) peer review but higher than any internal review approach.

However, there is a rather large disadvantage as of today. Not only does no such certification and/or agency exist for the specific problem of research data quality in materials science but also the cost must be considered as significantly higher than the other presented approaches given the evidence of ISO9001 certification suggests. As long as these factors are not changed, a successful utilization of such an approach is highly unlikely.

**4.2.5. External review (social web, e.g., "likes" - facebook; "comments/stars" - amazon)**

Another possible approach based on external, independent review is to utilize the increasing power of the possibilities the social web offers. Inspired by established systems like the online review system (stars) by amazon (www.amazon.com) for products, ratings (stars) for seller/buyer on ebay (www.ebay.com), the recommendation system on LinkedIn (www.linkedin.com) or the 'like' button on facebook (www.facebook.com). The system utilizes the power of many and allows even a dedicated rating of certain aspects of a data set comparable to the seller rating on ebay/amazon regarding quality of the product, timeliness of delivery and communication. This could be translated as independent aspects concerning the quality of research data sets e.g., based on the IQ dimensions presented above. Researchgate (www.researchgate.com) already offers the possibility to publish data sets (all disciplines and formats) and allows the community to discuss and up-/down-vote it. There are many different options of structuring such an approach. The reviews can be in the form of a written comment, where either everybody can be allowed to review the data or just reviewers who used the data in an accepted, reviewed publication (proof by citation) may be allowed to review. This would have the advantage that the reviewer can be assumed to have gained the needed knowledge of the data set during the preparation of the publication to review the data set. However, this might introduce another bias as the reviewer then might lean towards having a positive attitude towards the data quality from the beginning.

Overall, such an approach offers both advantages and disadvantages. Among the advantages are the low cost and minimal time and effort required from the data generating party. Disadvantages are the problem of preventing fraud, common standards/comparability and the acceptance within the community. As it is a very special domain, the design of the social review system can help to minimize certain disadvantages. For example, an obligatory registration with an institutional email address may help to legitimize the reviews authenticity. Furthermore, a mandatory and openly accessible personal profile including links of the reviewer / reviewers institution to the author / authors institution as well as a review history may help to uncover potential bias.

Such an approach, depending on the organization (see above), in general has a problem with its legitimacy and the number of potential reviewers willing to invest time. The amazon approach of using the power of the many in order to ensure the legitimacy of the merged reviews depends on a large number of reviewers. An approach leaning more to the written comment can be seen as a voluntary open peer review with all the advantages and disadvantages such an approach offers.

#### 4.2.6. Automated review (e.g., software semantics / statistical methods)

The last approach is based on an automated review based on e.g., software semantics or statistical methods. Such an approach offers the advantage of high comparability and standardization as well as being comparable low cost (Blumberg, Chaty & Kleinhof, 2011). Furthermore, the time and effort invested in the data generating party can be considered low. If implemented right, such an approach can even support the data generating process by alerting the parties involved when certain data pieces are missing or indicate incoherency. This is an important factor none of the other approaches offer (except maybe the first, having the author responsible) as they launch after the data set is already generated. Furthermore, an automated system is less likely to being exposed to fraud, even though it might trigger a certain preparation of the data once the audience is aware of the focus area of the algorithm. Overall, it can be considered very invulnerable of fraud and manipulation.

An automated system depends strongly on the programming. This determines if there is a bias involved and in the long run, the acceptance within the community. The possibility to include artificial intelligence (AI) and machine learning (ML) techniques would allow the system to learn actively from past reviews and increase the review quality over time. However, this may involve significant resources to develop and evaluate the to-be developed AI and/or ML systems depending on the requirements. On the other hand, even a relatively simple automated approach requiring little development resources, might improve the data set review and thus the data quality in combination with other approaches by automatically checking the data set for inconsistencies, missing values and statistical errors and providing a report to the (human) reviewers for a more detailed review.

The six approaches described in this section are evaluated in the following section against the previously presented aspects of what has to be considered when research data quality needs to be assured.

### 4.3. Assessment of suitable research data quality insurance approaches

In this section, the previously presented approaches are mapped to the identified requirements in form of a table (see Table 1). The assessment of suitability is conducted qualitatively based on the discussion in the previous section. It is conducted in form of an attribution of one (*) to five (*****) stars, one star meaning the approach does not address the requirement whereas five stars indicate high compliance with the requirement. For approach five, there are two ratings, one solely evaluating the impact on the described requirement when generally everybody is allowed to write a review, the other, including the additional stars in brackets, describes the evaluation result when the reviewers have to proof their suitability through usage of the data set in a publication. In the last column, the possible combinations with other approaches, which are assumed to have a positive effect on the impact, are indicated by the number of the other approaches.

Table 1.  Mapping of individual approach to identified requirements of methodology

| Requirements<br><br>Approach | Access to and knowledge of data & meta-data | Preventing fraud | Same standards / comparable | Cost | Needed time / effort | Starting point / Influence during process | Overall acceptance by research community | Positive effects of combination with following other approaches |
|---|---|---|---|---|---|---|---|---|
| *1. Author responsible* | ***** | * | ** | ***** | **** | ***** | ** | 2. / 3. / 4. / 5. / 6. |
| *2. Internal review* (e.g., superior) | **** | ** | ** | **** | *** | *** | *** | 1. / 3. / 4. / 5. / 6. |
| *3. External review* (peers) | ** | *** | *** | *** | * | * | **** | 1. / 2. / 6. |
| *4. External review* (3rd party) | *** | ***** | ***** | * | ** | ** | **** | 1. / 2. / 6. |
| *5. External review* (social web) | * (**) | **** | *** | *** | *** | * | * (**) | 1. / 2. / 6. |
| *6. Automated review* | *** | ***** | ***** | ***** | ***** | ***** | *** | 1. / 2. / 3. / 4. / 5. / 6. |

## 5. DEVELOPMENT OF SYSTEMATIC METHODOLOGY

In this section, the previous (mostly theoretical) discussion is included in a practical methodology, which is implemented in a developed materials science research data infrastructure. Therefore, the chosen approaches and their implementation are described including our fundamental motivation.

### 5.1. Systematic methodology and selected approaches

As all of the previously presented approaches have their specific strength and weaknesses concerning the overall goal of ensuring data quality, thus support the reuse of experimental research data in materials science in a different way. Therefore, in order to create a comprehensive and sustainable methodology, a combination of different approaches is chosen for the systematic approach presented in the following.

Table 1 indicates that a combination of different approaches is possible and might have a positive impact on the overall performance of the methodology. Whereas the approaches dealing with the different forms of external review (3. / 4. / 5.) are not positively correlated to each other, each individual one can be combined with a set of the other three approaches (1. / 2. / 6.). Overall, the basic principle of the methodology is that the author is responsible for the experimental design, generation of the research data set and its quality in the first place. This is necessary as in materials science, the experimental layout has a great influence on the data and the knowledge (access to knowledge) is necessary to set up the data set according to standards. This is a basic requirement for future continuous exploitation of the research data. All other approaches are additionally installed to support this basic responsibility of the author.

The project manager or the advisor/supervisor of the author, depending on the organizational structure, is additionally responsible for a first quality check, especially for the compliance with established quality management processes within the organization. This is an important step towards a standardized and therefore comparable creation of experimental research data in materials science.

From the three variations of external review, the social web approach (5.) is chosen within this methodology. The reasons are that in the moment, no established external third party exists which could be utilized to act as a reviewer according to approach No. 4. and the additional cost of such an approach would hinder the acceptance within the field even further. The third approach however, external peer review, is partly combined with the chosen variation of the fifth approach. At the beginning, the review of the data set is limited to researchers who have worked with the data set, which has to be proven through citations and their unique profile, indication of their position, relation to the author and review history. Thus, the reviewers can be assumed to have in-depth knowledge of the data set and, through the transparency, the acceptance of the review within the community can be assured.

The approach dealing with automated quality assurance is a very important one, especially considering its future development possibilities. In a first step, the developed methodology incorporates the following automated quality assurance features supporting the author in creating a high quality research data set:

- Automatic check calculation of Key Performance Indicators (KPIs) (e.g., coefficient of elasticity) based on entered (manually and/or automatically) values
- Automated guidance of author to enter the required values for standardized experiments so human error is minimized concerning missing values
- Automated check for unusual values based on existing data sets and standardized calibration values allowing the author to double check the experimental parameters directly
- Automatically connecting the data set to environmental parameters (e.g., temperature, machine number, etc.) to create important meta-data

The methodology, incorporating different approaches to assure research data quality is implemented in an information infrastructure for materials science research data and the supporting organizational processes are designed as well in order to support the authors in the creation of their data sets. In the following section, the methodology's application within the system is briefly described before the methodology is evaluated according to the IQ dimensions in the section thereafter.

## 5.2. Discussion of results

In this section, the presented methodology and its incorporating approaches to assure research data quality for potential research in the materials science domain is evaluated. In the following table (Table 2), the 15 Information Quality (IQ) dimensions (Rohweder et al., 2011) presented before are utilized to structure the evaluation. The (expected) impact of the methodology is mapped to each individual dimension and judged by its ability to support the effort.

Table 1. Evaluation of developed research data quality assurance methodology based on the 15 IQ dimensions (Rohweder et al., 2011)

| IQ dimensions (Rohweder et al., 2011) | | Addressed by developed research data quality assurance methodology |
|---|---|---|
| System support | *Accessibility* | Quality of data set is focused on and therefore resistance to make it publically available may be minimized<br>Data sets will be uniquely identified through DOI number |
| | *Ease of manipulation* | Methodology supports the usage of data for different purposes by putting a focus of the continuous exploitation ability of the data set from the beginning |
| Inherent | *Reputation* | By combining the above mentioned approaches, the acceptance within the community is assumed to be high |
| | *Free of error* | Through the combination of the quality assurance approaches, the chances of potential errors within a data set are be minimized |
| | *Objectivity* | By combining automated approaches and external reviewers (low bias) with internal quality control, the objectivity of the results may increase |
| | *Believability* | Implementing not only quality assurance approaches but also organizational procedures addressing and supporting generation of data in experiments may increase credibility |
| Representation | *Understandability* | A main goal of the information infrastructure developed is the semantic annotation of research data which increases the understandability of the presented data especially for external users |
| | *Concise representation* | The data is saved following standardized rules in an open format |
| | *Consistent representation* | Through semantic annotation and standardized processes the consistency of representation can be assured |
| | *Interpretability* | Semantic annotation assures interpretability of the data and meta-data |
| Purpose-dependent | *Timeliness* | Different starting points of the incorporated approaches assure the timeliness of quality assurance when it is needed |
| | *Value-added* | Every reuse and the continuous exploitation increases the value add of the data sets as it involves a fraction of the cost needed to generate it again by experiments |
| | *Completeness* | Both human and automated approaches focus on the assurance of completeness of the data set |
| | *Appropriate amount of data* | Processes and automated support approaches are based on existing standards and experience. Following credo 'as much as necessary, as little as possible' |
| | *Relevancy* | Both automated and human approaches focus on creating data sets including all necessary data for continuous exploitation of the data set in different applications |

As can be seen in the previous table, all IQ dimensions are addressed by the developed methodology. Furthermore, the performance in each IQ dimension is expected to improve after the methodology is utilized and implemented in the system. However, there are certain limitations and challenges, which may hinder the successful implementation of the methodology. First, the existing variety of experiments and new experimental designs are challenging when it comes to the development of standardized processes. This directly influences the possible future development of an automated quality assurance approach. Within the presented methodology, the semantic annotation allows a continuous extension of the system. Thus, additional experimental designs can be incorporated in the existing system. Second, the involvement of external reviewers through the social web features may have to reach a critical mass and this will take a certain amount of time. This is especially true due to the (at the beginning) agreed selection of reviewers by usage of the data set provable by citations. Until a publication using a data set is cited, at least six months can be calculated. A third major point is the question, how to motivate researchers to participate in sharing their data especially when a culture of sharing is not yet established in the community. This cannot be solved solely by a technical solution, but the usability and added values such a system presents, might help in the process. Whereas a system with immanent problems and an unclear structure will most likely hinder the development towards a sharing culture. Other limitations and challenges are more of a strategic relevance. Among those is the important question of how data and experiments of industrial projects with a signed confidentiality agreement may be incorporated in such systems.

## 6. CONCLUSION AND OUTLOOK

In this paper, first the specific nature of experimental research data in the domain of materials science was established. Following, the current state of the art concerning information and data quality was presented, including the information quality (IQ) dimensions (Rohweder et al., 2011). After looking into the specific meaning of data quality for materials science experimental research data, requirements for research data quality assurance approaches for information infrastructures were derived. Based on the findings, six data quality assurance approaches were developed and subsequently evaluated regarding their compliance with the previously developed requirements. Afterwards, a systematic methodology incorporating a combination of suitable approaches based on the previous discussion was presented including case based practical experience. Finally, the methodology and its indicated results were evaluated in a structured way using the implication on the information quality dimensions.

The findings indicate that experimental research data in the materials science domain represents specific challenges concerning the diverse nature and high-dimensionality of the resulting data and meta-data. This is caused by a variety of factors like e.g., the large number of tests, process parameters, machines, etc. adding to the inherit complexity are the many different data formats (data format (e.g., photos, graphs, plots, data all analog and/or digital) and available standards (e.g., DIN, ISO, EN, ASTM).

Overall, no individual approach on its won was found to ensure the experimental research data quality according to the derived requirements. However, by combining suitable approaches within the methodology, the requirements may be fulfilled to a satisfying degree. Subsequently, the acceptance within the community and, in the long run, the availability of high quality materials science research data sets for continuous exploitation may increase. Hence, the return on investment for funding agencies might increase accordingly when the generated data can be exploited for various research objectives in different institutions.

However, there are several areas where further research is necessary. First of all, the practical incorporation of the methodology and the continuous development including feedback from users especially of the automated approach need to be investigated further. Secondly, the social web features and its acceptance by the community, as well as measures to actively promote the usage present further research opportunities. Lastly, the impact on the data quality has to be critically evaluated after the system is fully functional and running based on a comparison of existing materials science experimental data sets before and after implementation of the methodology.

## ACKNOWLEDGEMENT

This work was partly funded by the "Deutsche Forschungsgemeinschaft" via the project "Informationssystem für werkstoffwissenschaftliche Forschungsdaten" and the German Academic Exchange Service (DAAD), Germany, through a Ph.D. fellowship (DAAD Jahresstipendium für Doktoranden). The authors would like to extend their gratitude to the funding parties for their generous support.

## AUTHORS

**Thorsten Wuest** earned a Master's from University of Bremen and a Master's from Auckland University of Technology. He has worked at BIBA, Bremen since 2009, while concurrently pursuing his dissertation project. In 2013/2014 he was furthering his research for ten months at the University of Southern California (USC). He is currently working among other things in the areas of knowledge, information and data management as well as closed-loop, item-level PLM with a focus on manufacturing.

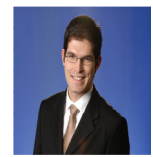

**Rainer Tinscher,** born in 1969, studied from 1990 to 1995 production engineering at the University of Bremen and earned his PhD in 2003 in the field of materials science. Since 2000 he has been working in the Stiftung Institut für Werkstofftechnik Bremen (IWT Bremen), where he works today as chief engineer.

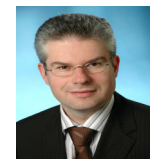

**Robert Porzel** earned degrees in general linguistics, computational linguistics and computer science. At the University of Bremen, Robert Porzel started to work on contextual computing for autonomous collaborative systems in the DFG-funded special research program "Autonmous Logistic Processes" (SFB 637). He initiated and chaired the biannual ScaNaLU workshop series on scalable natural language understanding systems (2002, 2004 and 2006) and is member of the steering board of the Interdisciplinary College (IK) - an international spring school on cognitive science, neurosciences and AI.

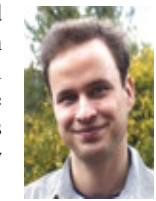

**Klaus-Dieter Thoben** studied mechanical engineering with a focus on product development at the TU Braunschweig, Germany. He received Ph.D. in the field of CAD/PDM applications. In 1989 he joined BIBA as Head of the Department of Computer Aided Design, Planning and Manufacturing. In 2001 he received a state doctorate (Habilitation) including a venia legendi for the domain production systems / production systematic. He was appointed a full professor at the University of Bremen, Germany in 2002 and currently is head of the "Institute of integrated Product Development" at the production engineering faculty. Additionally, he was named director at BIBA in the same year.

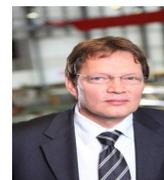